\documentclass[runningheads]{svmult}

\usepackage{makeidx}   
\usepackage{graphicx}  
\usepackage{subeqnar}  
\usepackage{multicol}  
\usepackage{physprbb}  
\makeindex             

\def\Reff{R_{\mathrm e}}

\def\csiv{{\vec{x}'}}
\def\nuv{{\vec{v}'}}
\def\Omegv{\vec{\Omega}}
\def\xv{\vec{x}}

\def\csici{x'_i}

\def\ddcsiv{\ddot{\csiv}}

\def\dOmegv{\dot{\Omegv}}
\def\ddxv{\ddot{\xv}}

\def\Rot{{\cal R}}
\def\RotT{\Rot^{\rm T}}

\def\phiM{\varphi_{\rm M}}
\def\theM{\vartheta_{\rm M}}
\def\psiM{\psi_{\rm M}}

\def\omegphi{\omega_{\varphi}}
\def\omegthe{\omega_{\vartheta}}
\def\omegpsi{\omega_{\psi}}

\def\dphi{\dot\varphi}
\def\dtheta{\dot\vartheta}
\def\dpsi{\dot\psi}

\def\ddphi{\ddot\varphi}
\def\ddtheta{\ddot\vartheta}
\def\ddpsi{\ddot\psi}

\def\Tc{\vec{T}}
\def\Ti{T_i}
\def\Tj{T_j}
\def\Tu{T_1}
\def\Td{T_2}
\def\Tt{T_3}

\def\DT{\Delta T}
\def\DTij{\DT_{ij}}
\def\DTdu{\DT_{21}}
\def\DTtu{\DT_{31}}
\def\DTtd{\DT_{32}}
\def\DTud{\DT_{12}}
\def\DTut{\DT_{13}}
\def\DTdt{\DT_{23}}

\def\Ii{I_i}
\def\Iu{I_1}
\def\Id{I_2}
\def\It{I_3}
\def\DI{\Delta I}
\def\DIdu{\DI_{21}}
\def\DItu{\DI_{31}}
\def\DItd{\DI_{32}}

\def\phig{\phi_{\mathrm g}}
\def\rhog{\rho_{\mathrm g}}
\def\Mg{M_{\mathrm g}}
\def\Mgn{M_{\mathrm g,11}}
\def\hg{h_{\mathrm g}}

\def\alphau{\alpha_1}
\def\alphad{\alpha_2}
\def\alphat{\alpha_3}
\def\alphai{\alpha_i}
\def\alphaj{\alpha_j}
\def\alphak{\alpha_k}
\def\alphaun{\alpha_{1,1}}

\def\acu{a_1}
\def\acd{a_2}
\def\act{a_3}

\def\acun{a_{1,250}}

\def\Porb{P_{\mathrm orb}}
\def\Pdyn{P_{\mathrm dyn}}

\def\msun{M_{\odot}}

\def\rhoc{\rho_{\mathrm c}}
\def\rhocz{\rho_{\mathrm c,0}}
\def\Mc{M_{\mathrm c}}

\def\sigv{\sigma_{\mathrm V}}
\def\sigvn{\sigma_{\mathrm V,1000}}
\def\sigvd{\sigv^2}

\def\Pphi{P_{\varphi}}
\def\Pthe{P_{\vartheta}}
\def\Ppsi{P_{\psi}}

\def\wci{w_i}
\def\wcu{w_1}
\def\wcd{w_2}
\def\wct{w_3}

\begin{document}
\title*{Collisionless evaporation from cluster elliptical galaxies}

\author{V. Muccione\inst{1} \and L. Ciotti\inst{2}}

\institute{Geneva Observatory, 51. ch. des Maillettes, 
           1290 Sauverny, Switzerland
\and       Dipartimento di Astronomia, Universit\`a di Bologna,
           via Ranzani 1, 40127 Bologna, Italy}

\maketitle              

\begin{abstract}
We describe a particular aspect of the effects of the parent cluster
tidal field (CTF) on stellar orbits inside cluster Elliptical galaxies
(Es). In particular we discuss, with the aid of a simple numerical
model, the possibility that {\it collisionless stellar evaporation}
from elliptical galaxies is an effective mechanism for the production
of the recently discovered intracluster stellar populations (ISP). A
preliminary investigation, based on very idealized galaxy density
profiles (Ferrers density distributions), showed that over an Hubble
time, the amount of stars lost by a representative galaxy may sum up
to the 10\% of the initial galaxy mass, a fraction in interesting
agreement with observational data.  The effectiveness of this
mechanism is due to the fact that the galaxy oscillation periods near
equilibrium configurations in the CTF are comparable to stellar
orbital times in the external galaxy regions. Here we extend our
previous study to more realistic galaxy density profiles, in
particular by adopting a triaxial Hernquist model.
\end{abstract}

\section{Introduction}

Observational evidences of an Intracluster Stellar Population (ISP)
are mainly based on the identification of {\it intergalactic}
planetary nebulae and red giant stars (see, e.g.,
~\cite{journ1},\cite{journ2},\cite{journ3},\cite{journ4},\cite{journ5}).
Overall, the data suggest that approximately 10\% (or even more) of
the stellar mass of clusters is contributed by the ISP ~\cite{journ6}.
The usual scenario assumed to explain the finding above is that
gravitational interactions between cluster galaxies, and interactions
between the galaxies with the gravitational field of the cluster, lead
to a substantial stripping of stars from the galaxies themselves.

Here, supported by a curious coincidence, namely by the fact that {\it
the characteristic times of oscillation of a galaxy around its
equilibrium position in the cluster tidal field (CTF) are of the same
order of magnitude of the stellar orbital periods in the external part
of the galaxy itself}, we explore the effects of interaction between
stellar orbits inside the galaxies and the CTF.  In fact, based on the
observational evidence that the major axis of cluster Es seems to be
preferentially oriented toward the cluster center, N-body simulations
showed that model galaxies tend to align, as observed, reacting to the
CTF as rigid bodies ~\cite{CD94} .  By assuming this idealized scenario, 
a stability analysis then showed that this configuration
is of stable equilibrium, and allowed to calculate the oscillation
periods in the linearized regime ~\cite{CG98}.  In particular,
oscillations around two stable equilibrium configurations have been
considered, namely: 1) when the center of mass of the galaxy is at
rest at center of a triaxial cluster, and the galaxy inertia ellipsoid
is aligned with the CTF principal directions, and 2) when the galaxy
center of mass is placed on a circular orbit in a spherical cluster,
and the galaxy major axis points toward the galaxy center while the
galaxy minor axis is perpendicular to the orbital plane.

Here, prompted by these observational and theoretical considerations,
we extend a very preliminary study of the problem~\cite{MC03}, by
evolving stellar orbits in a more realistic galaxy density profile:
for simplicity we restrict to case 1) above, while the full
exploration of the parameter space, together with a complete
discussion of case 2), will be given elsewhere \cite{CM03}. It is
clear, however, that both cases are rather exceptional. Most cluster
galaxies neither rest in the cluster center nor move on circular
orbits, but they move on elongated orbits with very different
pericentric and apocentric distances from the cluster's center; in a
triaxial cluster many orbits are boxes and some orbits can be
chaotic. These latter cases can be properly investigated only by
direct numerical simulation of the stellar motions inside the
galaxies, coupled with the numerical integration of the equations of
the motion of the galaxies themselves.

\section{The physical background}

Without loss of generality we assume that in the (inertial) Cartesian
coordinate system $C$, with the origin on the cluster center, the CTF tensor
$\Tc$ is in diagonal form, with components $\Ti$ $(i=1,2,3)$.  By
using three successive, counterclockwise rotations ($\varphi$ around
$x$ axis, $\vartheta$ around $y'$ axis and $\psi$ around $z''$ axis),
the linearized equations of the motion for the galaxy near the equilibrium
configuration can be written as
\begin{equation}
\ddphi =
{\DTtd\DItd\over\Iu}\varphi, \qquad \ddtheta =
{\DTtu\DItu\over\Id}\vartheta,\qquad \ddpsi =
{\DTdu\DIdu\over\It}\psi,
\label{eqn1}
\end{equation}
where $\Delta T$ is the antisymmetric tensor of components
$\DTij \equiv \Ti-\Tj$, and $\Ii$ are the principal components of the
galaxy inertia tensor.  In addition, let us also assume that
$\Tu\geq\Td\geq\Tt$ and $\Iu\leq\Id\leq\It$, i.e., that $\DTtd, \DTtu$
and $\DTdu$ are all less or equal to zero (see, e.g.,
~\cite{CG98},~\cite{CM03}). Thus, the equilibrium position associated
with eq. (1) is {\it linearly stable}, and its solution is
\begin{equation}
\varphi   =\phiM \cos (\omegphi t), \quad 
\vartheta =\theM \cos (\omegthe t), \quad
\psi      =\psiM \cos (\omegpsi t),
\label{eqn2}
\end{equation}
where
\begin{equation}
\omegphi = \sqrt{\DTdt\DItd\over\Iu},\quad  
\omegthe = \sqrt{\DTut\DItu\over\Id},\quad 
\omegpsi = \sqrt{\DTud\DIdu\over\It}.
\label{eqn3}
\end{equation}

For computational reasons the best reference system in which calculate
stellar orbits is the (non inertial) reference system $C'$ in which
the galaxy is at rest, and its inertia tensor is in diagonal form. The
equation of the motion for a star in $C'$ is
\begin{equation}
\ddcsiv = \Rot^{\rm T}\ddxv 
        -2\Omegv\wedge\nuv 
        -\dOmegv\wedge\csiv 
        -\Omegv\wedge (\Omegv\wedge\csiv ),
\label{eqn4}
\end{equation}
where $\xv=\Rot (\varphi,\vartheta,\psi)\csiv$, and
\begin{equation} 
\Omegv = (\dphi\cos\vartheta\cos\psi +\dtheta\sin\psi , 
          -\dphi\cos\vartheta\sin\psi +\dtheta\cos\psi ,
          \dphi\sin\vartheta +\dpsi).
\label{eqn5}
\end{equation}

In eq. (4)
\begin{equation}
\RotT\ddxv = -\nabla_{\csiv}\phig +(\RotT\Tc\Rot)\csiv,
\label{eqn6}
\end{equation}
where $\phig (\csiv)$ is the galactic gravitational potential,
$\nabla_{\csiv}$ is the gradient operator in $C'$, and we used the
tidal approximation to obtain the star acceleration due to the cluster
gravitational field.

\begin{section}{Galaxy and cluster models}

For simplicity we assume that the galaxy and cluster densities are
stratified on homeoids.  In particular, the galaxy density belongs to
ellipsoidal generalization of the widely used $\gamma$-models
(\cite{D93},\cite{T94}):
\begin{equation}
\rhog (m) = {\Mg\over \alphau\alphad\alphat}
{3-\gamma\over 4\pi} {1\over m^\gamma (1+m)^{4-\gamma}},
\label{eqn7}
\end{equation} 
where $\Mg$ is the total mass of the galaxy, $0\leq\gamma\leq 3$ and
\begin{equation}
m^2 =  \sum_{i=1}^3 {(\csici )^2\over\alphai^2},
\qquad \alphau\geq\alphad\geq\alphat.
\label{eqn8}
\end{equation}

The inertia tensor components of a generic homeoidal density
distribution (in the natural reference system adopted in eq. (8)), are
given by
\begin{equation}
\Ii={4\pi\over 3}\alphau\alphad\alphat (\alphaj^2+\alphak^2)\hg,
\label{eqn9}
\end{equation}
where $\hg =\int_0^{\infty}\rhog (m)m^4 dm$, and so
$\Iu\leq\Id\leq\It$.  Note that, from eqs. (3) and (9) it results that
the frequencies for homeoidal stratifications {\it do not depend on
the specific density distribution assumed}, but only on the quantities
$(\alphau ,\alphad ,\alphat)$. We also introduce the two ellipticities
\begin{equation}
{\alphad\over\alphau}\equiv 1 -\epsilon, \quad\quad
{\alphat\over\alphau}\equiv 1 -\eta,
\label{eqn10}
\end{equation}
where $\epsilon\leq\eta\leq 0.7$.

\begin{figure}
\begin{center}
\includegraphics[width=.5\textwidth]{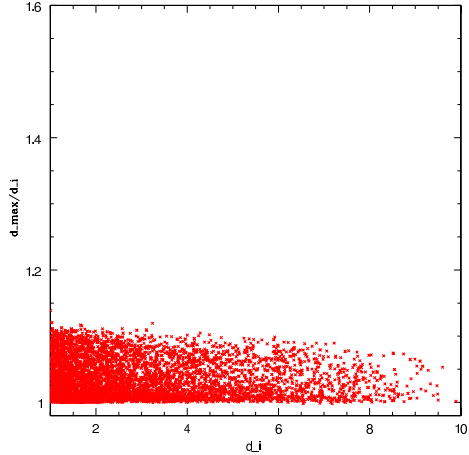}
\end{center}
\caption[]{Distribution of the $d_{\rm max}/d_{\rm i}$ ratio
vs. $d_{\rm i}/\alphau$ after an Hubble time for the model galaxy at
rest. $d_{\rm i}$ is the initial distance of the star from the galaxy
center, while $d_{\rm max}$ is the maximum distance from the galaxy
center reached during the simulation.}
\label{eps1}
\end{figure}

A rough estimate of {\it characteristic stellar orbital times} inside
$m$ is given by $\Porb (m)\simeq 4\Pdyn (m) = \sqrt{3\pi
/G\overline{\rhog} (m)}$, where $\overline{\rhog}(m)$ is the mean
galaxy density inside $m$. We thus obtain 
\begin{equation}
\Porb (m)\simeq 9.35\times 10^6 
                \sqrt{{\alphaun^3 (1-\epsilon)(1-\eta)\over\Mgn}}
                m^{\gamma/2}(1+m)^{({3-\gamma})/2} \quad \;{\rm yrs}, 
\label{eqn11}
\end{equation}
where $\Mgn$ is the galaxy mass normalized to $10^{11}\msun$,
$\alphaun$ is the galaxy ``core'' major axis in kpc units (for the
spherically symmetric $\gamma =1$ Hernquist model~\cite{H90},
$\Reff\simeq 1.8 \alphau$); thus, in the outskirts of normal galaxies
orbital times well exceed $10^8$ or even $10^9$ yrs.  For the
cluster density profile we assume
\begin{equation}
\rhoc (m) = {\rhocz\over (1+m^2)^2},
\label{eqn12}
\end{equation}    
where $m$ is given by an identity similar to eq. (8), with
$\acu\geq\acd\geq\act$, and, in analogy with eq. (10) we define
$\acd/\acu\equiv 1-\mu$ and $\act/\acu\equiv 1-\nu$, with
$\mu\leq\nu\leq 1$.  It can be shown (see, e.g.,
~\cite{CG98},\cite{CM03}) that the CTF components at the center
of a non-singular homeoidal distribution are given by
\begin{equation}
\Ti=-2\pi G\rhocz\wci (\mu,\nu),
\label{eqn13}
\end{equation}
where the dimensionless quantities $\wci$ are independent of the
specific density profile, $\wcu\leq\wcd\leq\wct$ for
$\acu\geq\acd\geq\act$, and so the conditions for stable equilibrium in
eq. (1) are fulfilled (\cite{CG98},\cite{CM03}).  The quantity
$\rhocz$ is not a well measured quantity in real clusters, and for its
determination we use the virial theorem, $\Mc\sigvd = -U$, where
$\sigvd$ is the virial velocity dispersion, that we assume to be
estimated by the observed velocity dispersion of galaxies in the
cluster. Thus, we can now compare the galactic oscillation periods:
\begin{subeqnarray}
\Pphi &=& {2\pi\over \omegphi} 
          \simeq {8.58\times 10^8\over\sqrt{(\nu -\mu)(\eta -\varepsilon)}}
          {\acun\over\sigvn} \; \mathrm{yrs} \; , \label{eqn14a}\\
\Pthe &=& {2\pi\over \omegthe}
          \simeq {8.58\times 10^8\over\sqrt{\nu\eta}}
          {\acun\over\sigvn} \; \mathrm{yrs}\; , \label{eqn14b}\\
\Ppsi &=& {2\pi\over\omegpsi} 
          \simeq {8.58\times 10^8\over\sqrt{\mu\varepsilon}}
          {\acun\over\sigvn} \; \mathrm{yrs}\; .\label{eqn14c}
\end{subeqnarray}
(for small galaxy and cluster flattenings, 
where $\acun = \acu/250$ kpc and $\sigvn =\sigv/10^3 \;
\mathrm{km}/\mathrm{s}$, \cite{CM03}) with the characteristic orbital times in
galaxies. Thus, from eq. (11) and eqs. (14abc), it follows that {\it
in the outer halo of giant Es, stellar orbital times can be of the
same order of magnitude as the oscillatory periods of the galaxies
themselves near their equilibrium position in the CTF}.  For example,
in a relatively small galaxy of $\Mgn =0.1$ and $\alphaun =1$,
$\Porb\simeq 1$ Gyr  at $m\simeq 10$ (i.e., at $\simeq 5\Reff$),
while for a galaxy with $\Mgn =1$ and $\alphaun =3$ the same orbital
time characterizes $m\simeq 7$ (i.e., $\simeq 3.5\Reff$).

\begin{figure}
\begin{center}
\includegraphics[width=.5\textwidth]{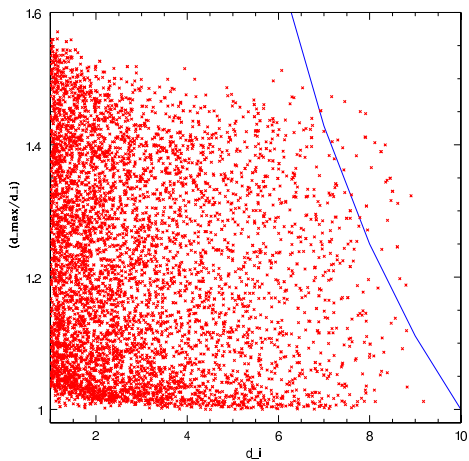}
\end{center}
\caption[]{Distribution of the $d_{\rm max}/d_{\rm i}$ ratio
vs. $d_{\rm i}/\alphau$ after an Hubble time for the same galaxy
model as in Fig. 1, when oscillating around its equilibrium position
in the CTF.}
\label{eps2}
\end{figure}

In order to understand the effects of the galaxy oscillations on the
stellar orbits, we performed a set of Monte-Carlo simulations, in
which we followed the evolution of $10^4$ - $10^5$ ``1-body problems''
over the Hubble time by integrating numerically eq. (4). At variance
with ~\cite{MC03}, where we used simple and easy-to-integrate Ferrers
density profiles, here we study orbital evolution in a more realistic
(but also more demanding from the numerical point of view) galaxy
density profile, namely a triaxial Hernquist model, obtained by
assuming $\gamma =1$ in eq. (7).  The gravitational potential inside
the galaxy density distribution, in a form suitable for the numerical
integration, was obtained by using an expansion technique useful in
case of small density flattenings (\cite{CM03},\cite{CB03}).  The
initial conditions are generated by using the Von Neumann rejection
method in phase-space (for details see ~\cite{CM03}): note that, at
variance with the analysis ~\cite{MC03}, now ``stars'' are
characterized by initial velocities that can be different from
zero. The code, a double-precision fortran code based on a Runge-Kutta
scheme, runs on GRAVITOR, the Geneva Observatory 132 processors
Beowulf cluster ({\bf
http://obswww.unige.ch/\~{}pfennige/gravitor/gravitor\_e.html}).  The
computation of $10^4$ orbits usually requires 2 hours when using 10
nodes.

\section{Preliminary results and conclusions}

We show here, as an illustrative case, the behavior of the ratio
$d_{\rm max}/d_{\rm i}$ as a function of $d_{\rm i}/\alphau$, for a
moderately flattened galaxy model ($\epsilon\simeq 0.2$ and
$\eta\simeq 0.3$), with $\Mg =10^{11}\,\msun$, semi-mayor axis
$\alphau =3$ kpc, and maximum oscillation angles equals to 0.1
rad. The cluster parameters are $\acun=\sigvn =1$, $\mu=0.2$,
$\nu=0.4$, and the total number of explored orbits is $N_{\rm
tot}=10^4$. In order to show the effect of oscillations, in the
following simulations we artificially eliminated the {\it direct}
contribution of the CTF, as given by the second term in the r.h.s. of
eq. (6).

In Fig. 1 we show the result of a first simulation in which the galaxy
is {\it not} oscillating: obviously, the ratio $d_{\rm max}/d_{\rm i}$
is in general (slightly) larger than unity, due to the initial
velocity of each star. In Fig. 2 we show the result for the same
galaxy model, when oscillating around the equilibrium position: the
effects of the galaxy oscillations are clearly visible as a global
``expansion'' of the galaxy. As a reference, the solid line indicates
the expansion ratio required to reach the representative distance of
$10\Reff$ from the galaxy center. Thus, it is clear that the galaxy
oscillations are certainly able to substantially modify the galaxy
density profile. In particular, it will be of interest the study of
the (more realistic) case in which the galaxy is in rotation around
the cluster center.  In this case we expect a different behavior of
stellar orbits as a function of the distance of the galaxy center of
mass from the cluster center: in fact, while inside the cluster core
the CTF is {\it compressive} (see, e.g., ~\cite{CD94},\cite{CG98}),
outside the CTF is {\it expansive} along the cluster radial direction,
and in this latter case its direct effect should increase the
expansive effect due to the galaxy oscillations. These cases are
discussed in detail in~\cite{CM03}.

\end{section}
\end{document}